\documentclass[final
  ,numberedheadings % uncomment this option for numbered sections
  ]
  {aipproc}

\layoutstyle{6x9}
\begin{document}

\title{Two-photon exclusive production of supersymmetric pairs at the LHC}

\classification{13.60.-r, 14.80.Ly}
\keywords      {Photon interactions, Forward scattering, Exclusive production, Kinematic constraints and reconstruction,Time-of-flight detectors}

\author{K. Piotrzkowski}{
  address={Center for Particle Physics and Phenomenology (CP3)\\
Universit\'{e} catholique de Louvain, 1348 Louvain-la-Neuve, Belgium}
}
\author{N. Schul}{
  address={Center for Particle Physics and Phenomenology (CP3)\\
Universit\'{e} catholique de Louvain, 1348 Louvain-la-Neuve, Belgium}
}

\begin{abstract}
The two-photon exclusive production of charged supersymmetric pairs at the \textsc{lhc} has a clean and unique signature - two very forward scattered protons and two opposite
charged leptons produced centraly. For low-mass 
\textsc{susy} scenarios, significant cross-sections are expected and background processes are well controlled. 
Measurement of the forward proton energies would allow for mass reconstruction of 
right-handed sleptons and the \textsc{lsp} with a few GeV resolution. Methods to reduce backgrounds at high luminosity resulting from accidental coincidences between
events in the central and forward detectors are discussed.
\end{abstract}

\maketitle

%%%%%%%%%%%%%%%%%%%%%%%%%%%%%%%%%%%%%%%%%%%%
%% MAINMATTER
%%%%%%%%%%%%%%%%%%%%%%%%%%%%%%%%%%%%%%%%%%%%

\section{High-Energy photon interactions at the \textsc{lhc}}
The $\gamma\gamma$ and $\gamma p$ interactions at the \textsc{lhc} offer interesting potential for the studies
of the Higgs boson, quartic gauge couplings and the top quark as well as for
searches of New Physics \cite{epj-opus}. In particular, the clean and striking experimental signatures of the exclusive 
processes $p p \rightarrow p X p$ (forward regions devoid of any hadronic activity apart from presence of two forward scattered protons)
are well suited for the search of new charged massive particles beyond the Standard Model \cite{zerwas,workshop}.

\section{Detection of exclusive \textsc{susy} pairs}
\begin{figure}[h]
\begin{tabular}{cc}
\includegraphics[width=0.26\textwidth]{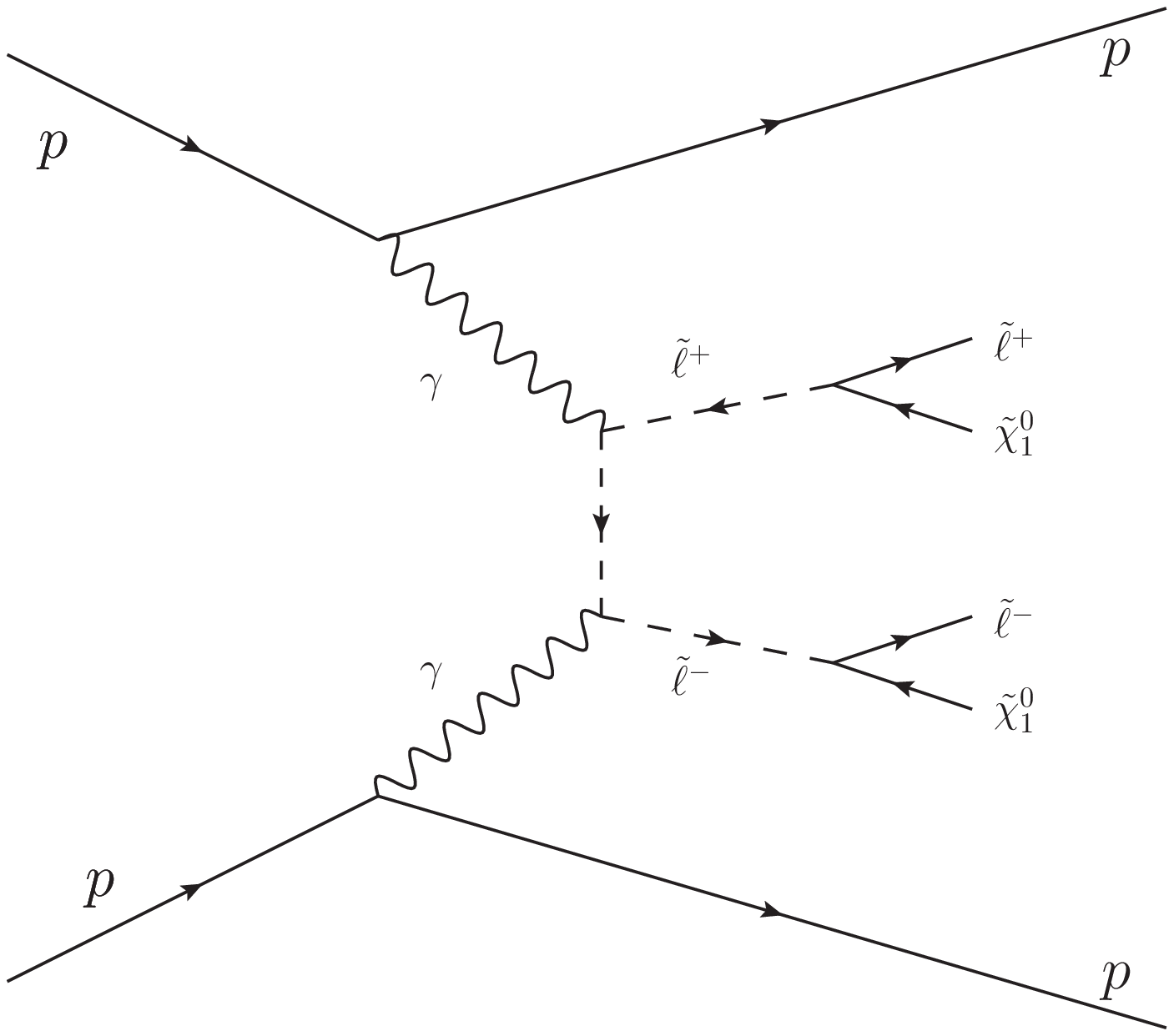} & \includegraphics[width=0.29\textwidth]{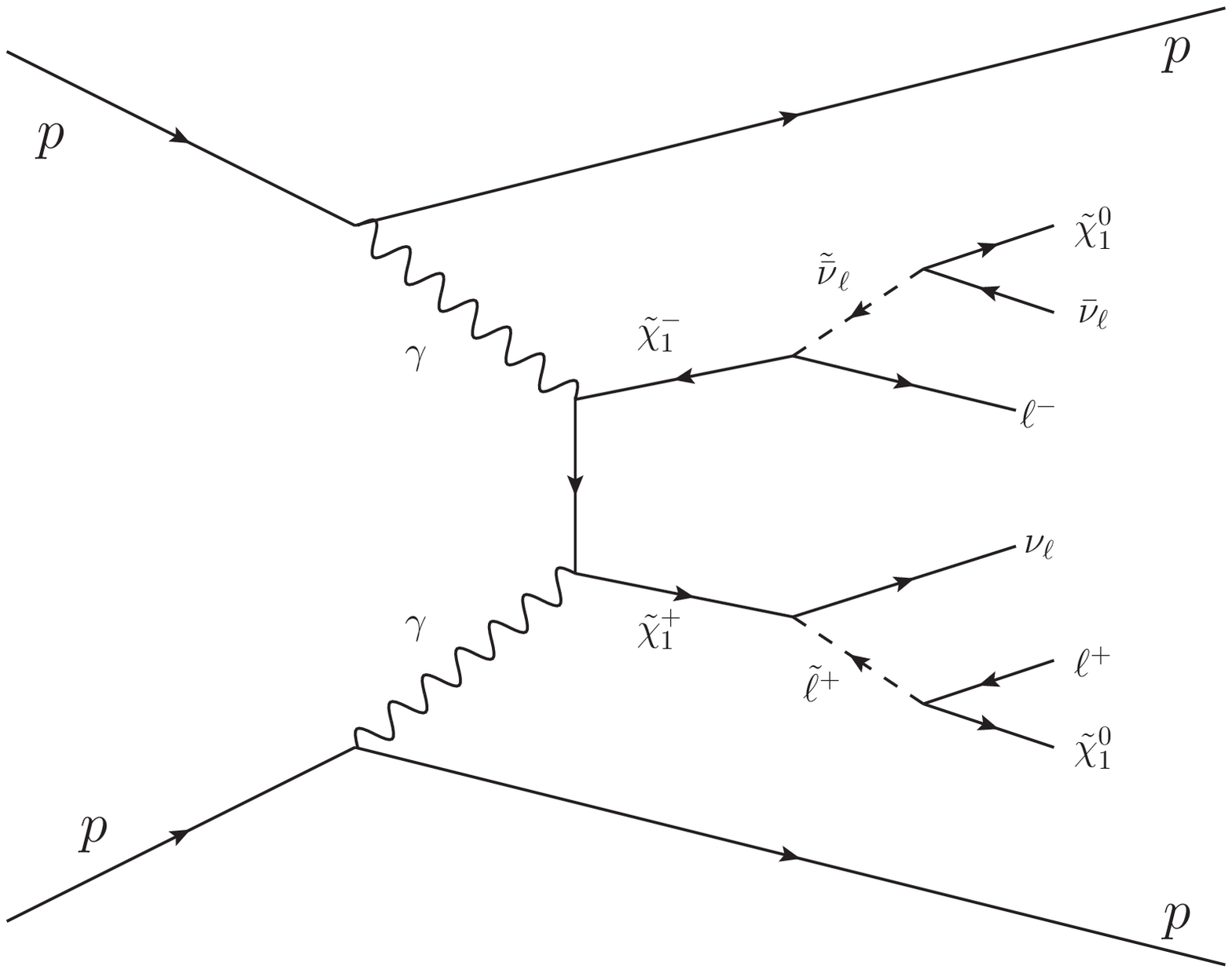} \\
\end{tabular}
\caption{Most relevant Feynman diagrams for $\gamma\gamma$ production of slepton pairs (left) and chargino pairs (right), both with two charged leptons in the final state.}
\label{fig:prod-decay}
\end{figure}

The two-photon pair production mechanism is simple and often results in simple final
states (as shown in Figure\ref{fig:prod-decay}), in contrast to many other cases at the \textsc{lhc} with usual complex
decay chain problems. In fact, the exclusive pair production cross-section
is given by the \textsc{susy} particle mass and its charge and spin \cite{epj-opus}. Only the final
states containing exactly two charged leptons have been assumed in this study as well
as the low mass point LM1 \cite{lm1} as a benchmark\footnote{All the details about 
the considered \textsc{susy} parameters and mass spectrum  can be found in \cite{workshop}.}. \\

Assuming a general multi-pupose \textsc{lhc} detector and a full-set of dedicated very forward detectors (\textsc{vfd}s) to tag $\gamma$-interaction \cite{piotr1,fp420}, the event selection requires:
\begin{itemize}
 \item  two isolated leptons of opposite charge detected in the central region assuming
\begin{equation}
p_T (\mu^{\pm})>7~\textrm{GeV}, \ \ \ p_T (e^{\pm})>10~\textrm{GeV}, \ \ \ |\eta(\ell^\pm)|<2.5
\label{eq:acceptance}
\end{equation}
 \item  two scattered protons detected in the \textsc{vfd}s assuming the tagged photon energy range \cite{hector}
\begin{equation}
20<E_{\gamma}<900~\textrm{GeV}
\label{eq:proton}
\end{equation}
 \item  large missing energy due to the escape of neutrinos and neutralinos\footnote{$\tilde{\chi}^0_1$ is taken as the Lightest \textsc{susy} Particle (\textsc{lsp}) in this model.},
 \item  lepton acoplanarity.
\end{itemize}
The only irreducible background process for such an event topology is the exclusive $\gamma\gamma$ production of $W$ pairs
with fully leptonic decay. The cross-sections after acceptance cuts (including same flavour requirement 
and rejection of $\tau$-lepton tagged events) 
are 0.56~fb and 1.46~fb for the signal and background respectively, 
as detailed in Table\ref{tab:signal}.

\begin{table}
~~~~~~~~~~~~~~~~~~~~~~~~~~~~~~~~~~~
\begin{tabular}{lccc}
\hline
\tablehead{1}{l}{b}{Processes} & 
\tablehead{1}{c}{b}{$\sigma$ [fb]} & 
\tablehead{1}{c}{b}{$\sigma_{acc}^{2p^+}$ [fb]} & 
\tablehead{1}{c}{b}{$\sigma_{ana}$ [fb]\tablenote{'$ana$' means $W_{miss}>194$~GeV, $W_{\gamma\gamma}>236$~GeV, $\Delta(\eta)<2.1$, $\Delta(R)<3.2$,
$P_T^{miss}>5$~GeV, $W_{lep} \not\in [87\textrm{~GeV};95\textrm{~GeV}]$}} \\
\hline
$\gamma\gamma \rightarrow \tilde{\ell}_R^+\tilde{\ell}_R^-$ & 0.798 & 0.445 & 0.357 \\ 
$\gamma\gamma \rightarrow \tilde{\ell}_L^+\tilde{\ell}_L^-$ & 0.183 & 0.093 & 0.073 \\ 
$\gamma\gamma \rightarrow \tilde{\tau}_i^+\tilde{\tau}_i^-$ & 0.604 & 0.001 & 0.001 \\ 
$\gamma\gamma \rightarrow \tilde{\chi}_i^+\tilde{\chi}_i^-$ & 0.642 & 0.021 & 0.015 \\ 
%$\gamma\gamma \rightarrow H^+H^-$                           & 0.004 & /     & /     \\ 
                                                            &       &       &       \\
$\gamma\gamma \rightarrow W^+W^-$                           & 108.5 & 1.463 & 0.168 \\ 
\hline
\end{tabular}
~~~~~~~~~~~~~~~~~~~~~~~~~~~~~~~~~~~
\caption{Cross-sections of exclusive signal processes for production ($\sigma$), after applying central and forward detectors acceptance cuts ($\sigma_{acc}^{2p^+}$) and after applying analysis cuts ($\sigma_{ana}$).}
\label{tab:signal}
\end{table}
%

%%%%%%%%%%%%%%%%%%%%%%%%%%%%%%%%%%%%%%%%%%%%
%% SAMPLE TABLE
%%
%% Shows the use of \tablehead and \tablenote
%% macros
%%%%%%%%%%%%%%%%%%%%%%%%%%%%%%%%%%%%%%%%%%%%

\section{Precise mass reconstruction}
\begin{figure}[h!]
\begin{tabular}{cc}
\includegraphics[width=0.45\textwidth]{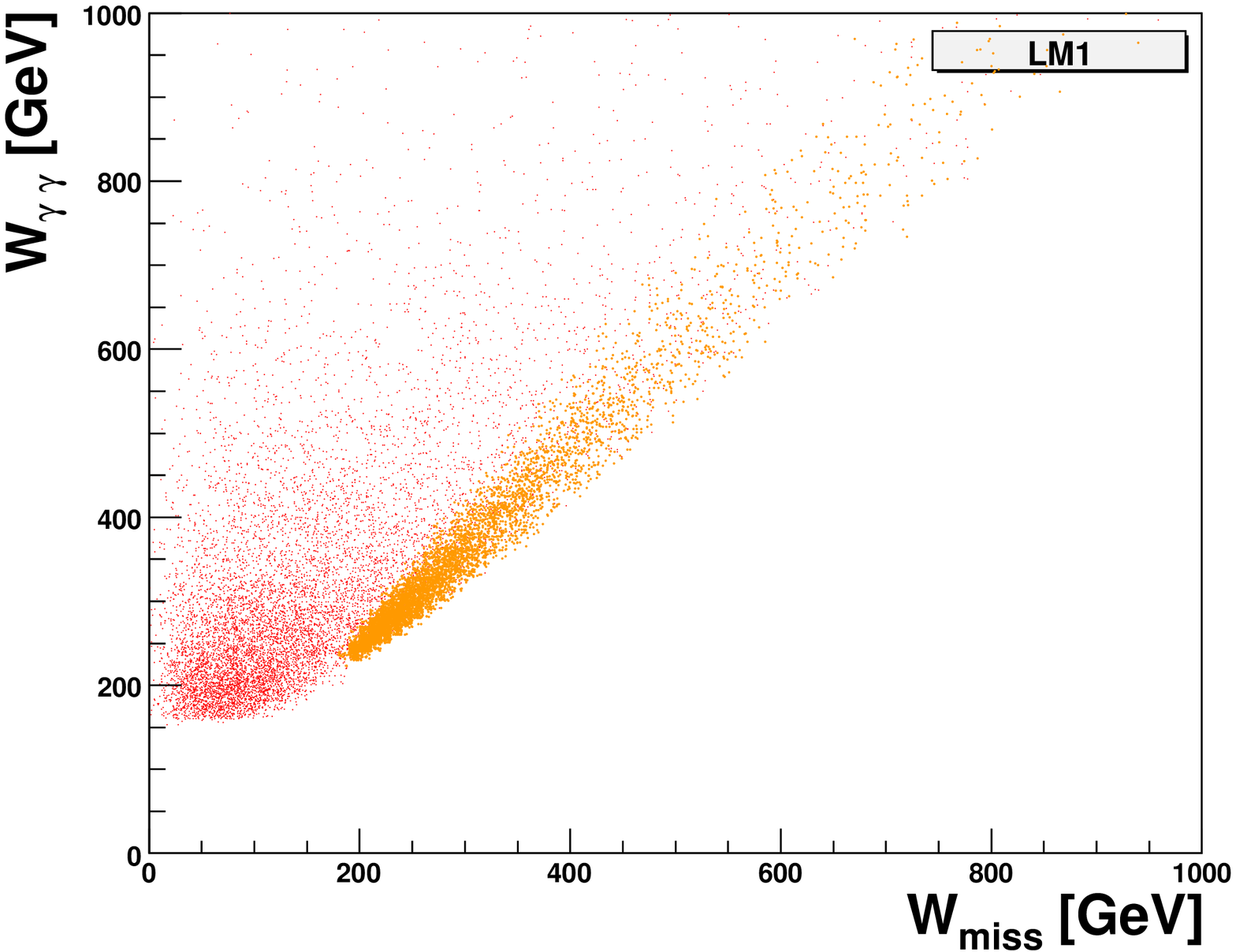} & \includegraphics[width=0.45\textwidth]{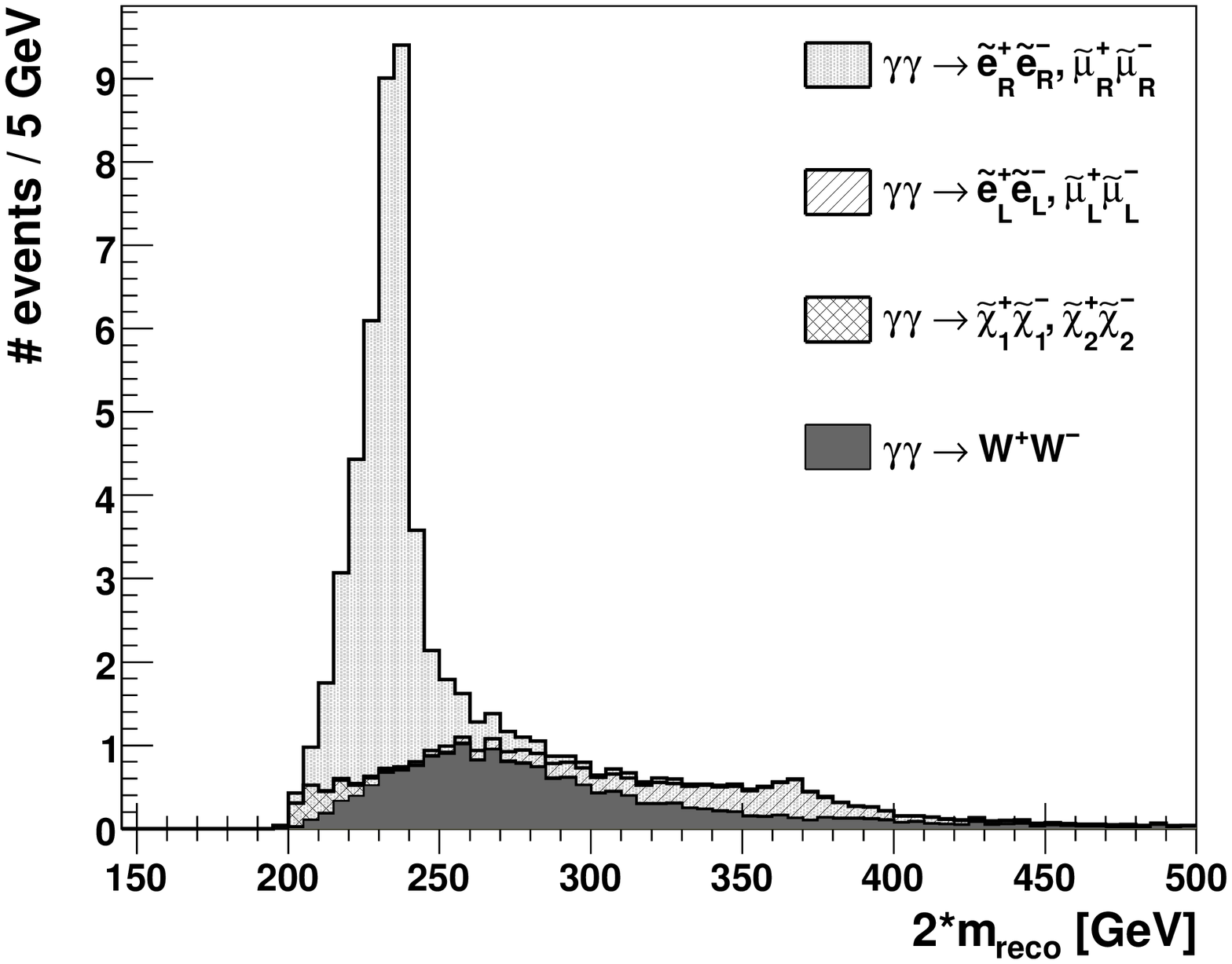}
\end{tabular}
\caption{Left: Scatter plots for the $\tilde{\mu}_R^+ \tilde{\mu}_R^-$ signal (light) and $WW$ background (dark) in the $W_{miss}$, $W_{\gamma\gamma}$ plane. Right: Cumulative distribution of the reconstructed mass $2 \times m_{reco}$ (\ref{eq:mreco}) for the integrated luminosity L = 100~fb$^{-1}$ assuming no pile-up conditions, after the analysis cuts.}\label{Fig:mreco}
\end{figure}

The detection of the two scattered forward protons and the associated measurement of the photon energies
allow for precise reconstruction of the two-photon 'initial conditions'.
\begin{figure}[h]
\begin{tabular}{cc}
\includegraphics[width=0.45\textwidth]{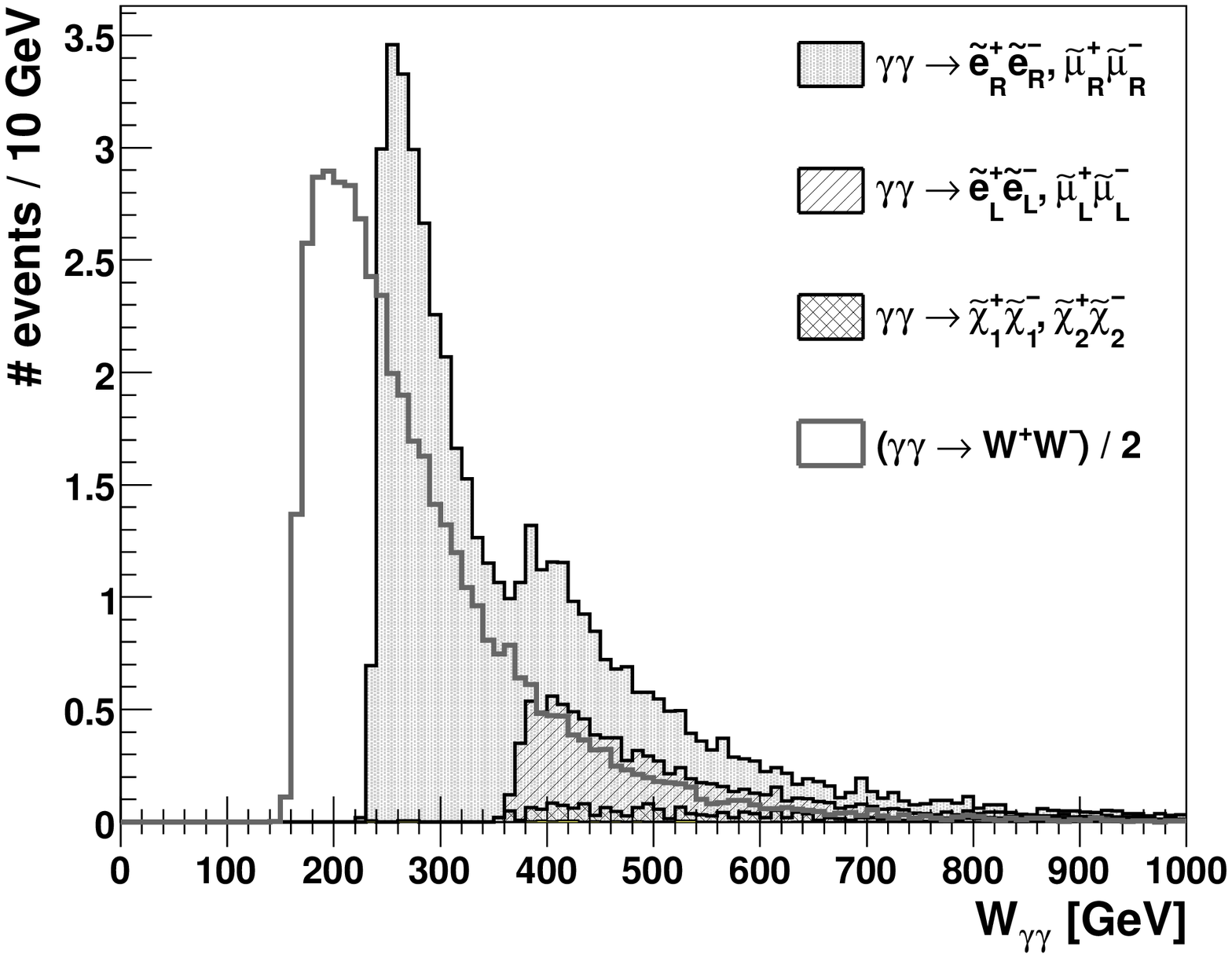} & \includegraphics[width=0.45\textwidth]{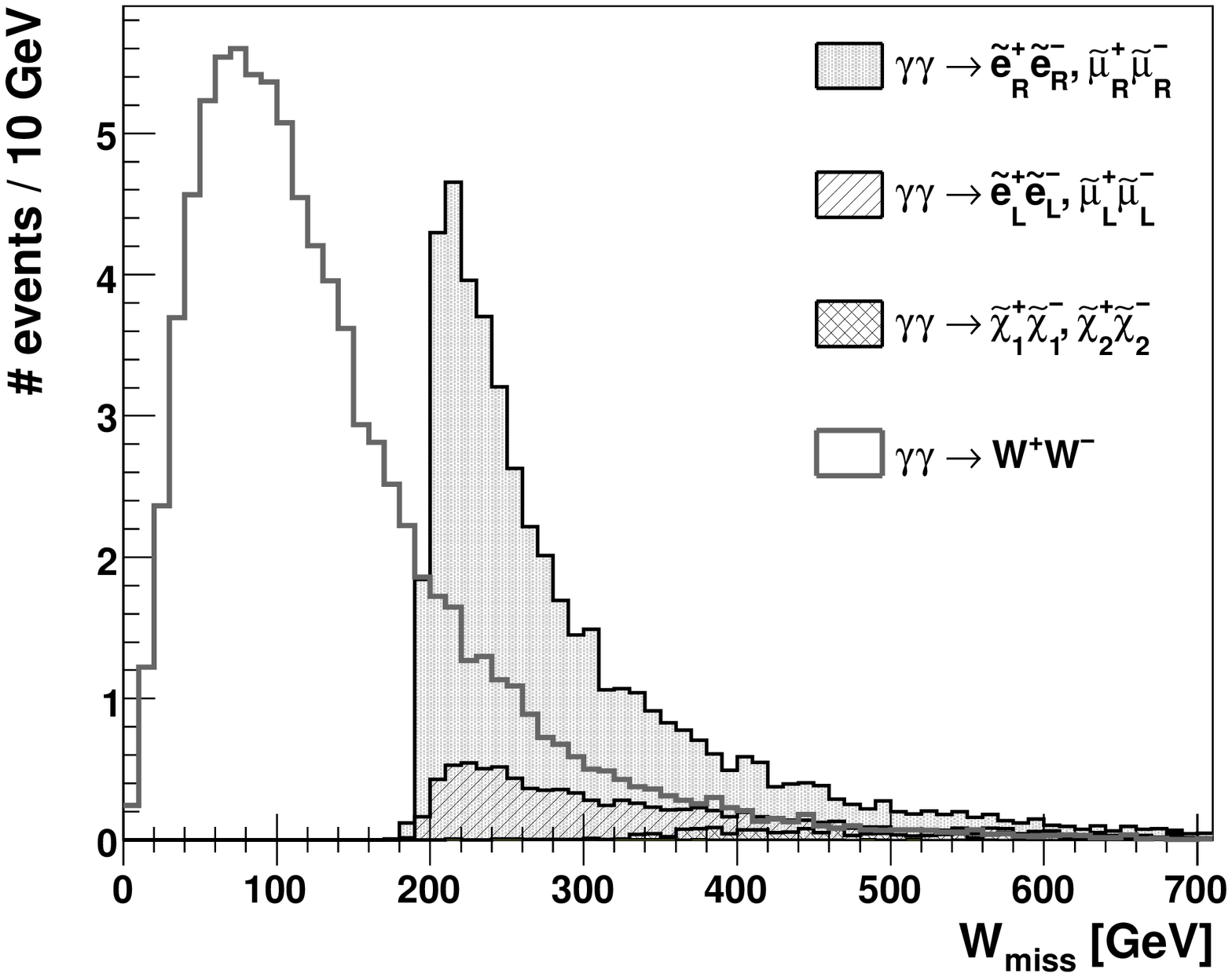} \\
\end{tabular}
\caption{Distributions of the two-photon invariant mass $W_{\gamma\gamma}$ (left) and the missing mass $W_{miss}$ (right) for the integrated luminosity
L = 100~fb$^{-1}$. The background distribution of $WW$ pairs is shown separately. All the distributions are obtained for
events passing the acceptance cuts (\ref{eq:acceptance}) and (\ref{eq:proton}) only, with 2 leptons of opposite
charge but same flavor.}
\label{fig:wgg-wmiss}
\end{figure}
The distributions of the two-photon invariant mass $W_{\gamma\gamma} = 2 \sqrt{E_{\gamma_1}E_{\gamma_2}}$ and of the missing mass $W_{miss}$ (reconstructed from
$E_{miss} = E_{\gamma_1}+E_{\gamma_2}-E_{\ell_1}-E_{\ell_2}$) are shown on Figure\ref{fig:wgg-wmiss}.
The $W_{\gamma\gamma}$ distribution reflects the \textsc{susy} mass spectrum with two peaks due to the production thresholds of $\tilde{\ell}_R$ pairs ($\simeq$250~GeV) and $\tilde{\ell}_L$ pairs ($\simeq$400~GeV). Similary the $W_{miss}$ distribution starts around twice the mass of the \textsc{lsp} for the signal, and around zero for 
the background. These distributions can be used to perform a mass edge study and extract the masses of $\tilde{\ell}_R^+$, 
$\tilde{\ell}_L^+$ and $\tilde{\chi}_1^{0}$. Furthermore, combination of both information allows to measure the mass of
light $\tilde{\mu}_R$ and $\tilde{e}_R$ using empirical quantity \cite{workshop}:
%\begin{flushleft}
%$(2m_{reco})^2 = W_{\gamma \gamma}^2 - ( [W_{miss}^2 - 4 m_{\tilde{\chi}_1^0}^2]^{1/2}$
%\end{flushleft}
%\begin{flushright} 
%$ + [W_{lep}^2 - 4 m_{lep}^2]^{1/2} )^2$
%\end{flushright}
\begin{equation}
(2 \times m_{reco})^2 = W_{\gamma \gamma}^2 - \left( [W_{miss}^2 - 4 m_{\tilde{\chi}_1^0}^2]^{1/2} + [W_{lep}^2 - 4 m_{lep}^2]^{1/2} \right)^2
\label{eq:mreco}
\end{equation}
The only input is this method is the \textsc{lsp} mass, which can be taken from the $W_{miss}$ edge study. 
The $m_{reco}$ distribution is shown in Figure\ref{Fig:mreco} after applying the analysis cuts (see Table\ref{tab:signal}) for the integrated luminosity of 100~fb$^{-1}$.
A narrow peak centered on $2 \times m_{reco} = $ 236~GeV = 2$\times$118~GeV demonstrates possibility of an event-by-event determination 
of the $\tilde{e}_R^\pm$ and $\tilde{\mu}_R^\pm$ mass with few GeV resolution.

\section{Accidental coincidence background}
To be sensitive to the fb-level cross-sections, such studies have to be performed
at the nominal \textsc{lhc} luminosity. The large event rates lead then to large backgrounds due
to accidental coincidences of dileptonic events in the central detector and two
forward protons detected in the \textsc{vfd}s, but not coming from the same vertex \cite{fp420,higgs}. The effect of multiple interactions per beam collision was simulated by superimposing over dileptonic inclusive events\footnote{We considered inclusive $Z/\gamma*$, $ZZ$ and $WW$ production.} $\langle N \rangle$ overlap events 
on average\footnote{$\langle N \rangle = 5.08$ for ${\cal L} = 2 \times 10^{33} cm^{-2} s^{-1}$ and
                    $\langle N \rangle = 25.38$ for ${\cal L} = 10^{34} cm^{-2} s^{-1}$},
and by distributing the extra vertices according to a $48.2$mm wide Gaussian \cite{white-correl}. The probability to detect two diffractive accidental protons per beam collision (one on each side of the interaction point) is expected to be on
average $1.2\%$ at low and $21.5\%$ at high luminosity. The associated background
cross-sections reach $2.4\times10^4$~fb and $4.5 \times 10^{5}$~fb respectively. \\

This background can be reduced using the tracks associated to the dilepton vertex,
which are not present in the exclusive events. One can therefore request no extra
track with $p_T>0.5$~GeV (where $100\%$ reconstruction efficiency is assumed) associated to the $\ell^+\ell^-$ vertex.
It provides a background reduction of $4500$. Further background
reduction can be made using precise time-of-flight proton detectors with a few
pico-second resolution, and by reconstruction of the event vertex position from the
proton timing \cite{fp420,gastof}.

%%%%%%%%%%%%%%%%%%%%%%%%%%%%%%%%%%%%%%%%%%%%%%%%
%% BACKMATTER
%%%%%%%%%%%%%%%%%%%%%%%%%%%%%%%%%%%%%%%%%%%%%%%%

\begin{theacknowledgments}
This work was supported by the Belgian Federal Office for Scientific, Technical and Cultural Affairs through the Interuniversity Attraction Pole P6/11.
\end{theacknowledgments}



\bibliographystyle{aipprocl} % if natbib is missing

%\bibliography{sample}

%  \typeout{******************************************}
%  \typeout{** Please run "bibtex \jobname" to optain}
%  \typeout{** the bibliography and then re-run LaTeX}
%  \typeout{** twice to fix the references!}
%  \typeout{******************************************}

%\end{document}

%%%%%%%%%%%%%%%%%%%%%%%%%%%%%%%%%%%%%%%%%%%
%% The following lines show an example how to produce a bibliography
%% without the help of the BibTeX program. This could be used instead
%% of the above.
%%%%%%%%%%%%%%%%%%%%%%%%%%%%%%%%%%%%%%%%%%%

%\endinput
\end{document}